# The Drain of Scientific Publishing

Fernanda Beigel[1,2], Dan Brockington[3,4,5]*, Paolo Crosetto[6], Gemma Derrick[7], Aileen Fyfe[8], Pablo Gomez Barreiro[9], Mark A. Hanson[10], Stefanie Haustein[11], Vincent Larivière[12,13,14], Christine Noe[15], Stephen Pinfield[16,17] and James Wilsdon[17]

[1] Instituto de Ciencias Humanas y Ambientales-Consejo Nacional de Investigaciones Científicas y Técnicas; Mendoza, Argentina.
[2] Centro de Estudios de la Circulación del Conocimiento-Universidad Nacional de Cuyo; Mendoza, Argentina.
[3] ICREA; Pg. Lluís Companys 23, Barcelona, Spain.
[4] Institut de Ciència i Tecnologia Ambientals de la Universitat Autònoma de Barcelona (ICTA-UAB); Barcelona, Spain.
[5] Department of Private Law, Universitat Autònoma de Barcelona; Barcelona, Spain.
[6] Univ. Grenoble Alpes, INRAE, CNRS, Grenoble INP, GAEL; Grenoble, France
[7] School of Education, University of Bristol; Bristol, UK.
[8] School of History, University of St Andrews; St Andrews, UK
[9] Department of Science Operations, Royal Botanic Gardens, Kew; Wakehurst, UK.
[10] Centre for Ecology and Conservation, University of Exeter; Penryn, Cornwall, UK
[11] Scholarly Communications Lab, School of Information Studies, University of Ottawa; Ottawa, Canada
[12] Consortium Érudit, Université de Montréal; Montréal, Quebec, Canada.
[13] Observatoire des Sciences et des Technologies, Université du Québec à Montréal; Montréal, Quebec, Canada.
[14] DSI-NRF Centre of Excellence in Scientometrics and Science, Technology and Innovation Policy, Stellenbosch University; Stellenbosch, South Africa.
[15] Department of Geography, University of Dar es Salaam; Dar es Salaam, Tanzania.
[16] School of Information, Journalism and Communication, University of Sheffield; Sheffield, UK
[17] Research on Research Institute (RoRI), Department of Science, Technology, Engineering & Public Policy, University College London; London, UK.
* Email: daniel.brockington@uab.cat

This is version 2 of the pre-printed paper.

ORCID IDs:
Beigel: 0000-0002-7996-9660; Brockington: 0000-0001-5692-0154; Crosetto: 0000-0002-9153-0159; Derrick: 0000-0001-5386-8653; Fyfe 0000-0002-6794-4140; Gomez Barreiro: 0000-0002-3140-3326; Hanson: 0000-0002-6125-3672; Haustein: 0000-0003-0157-1430; Larivière: 0000-0002-2733-0689; Noe: 0000-0002-3031-8074; Pinfield: 0000-0003-4696-764X; Wilsdon: 0000-0002-5395-5949

**Abstract:** The domination of scientific publishing in the Global North by major commercial publishers is harmful to science; we need the most powerful members of the research community – funders, governments and Universities – to lead the drive to re-communalise publishing to serve science not the market.

**Introduction**

Scholarly journals disseminate knowledge that advances our understanding of humanity, life and the universe. But they serve other purposes. They provide recognition and influence for researchers, their institutions and funders. Journals also earn revenue for commercial publishers, turning prestige into profit. These three purposes – knowledge, prestige and profit – are now poorly aligned.

In this article, we show that the relationships that have developed between researchers, their funders and commercial publishers are draining the research system, despite (sometimes even, due to) recent efforts to embrace Open Access publishing models. The drain is four-fold, depriving the research system of Money, Time, Trust and Control. In some languages, disciplines and regions, different publishing practices provide welcome alternatives. But as most researchers rely on publication in commercial journals to secure status and climb the prestige hierarchies, the drain continues to be at work.

In their early days, journals served small, dedicated communities of readers and often survived on philanthropy, altruism or institutional support (*1*). However, since the 1950s publications have become key tokens in the increasingly fierce competition for prestige. The number of publications worldwide increased exponentially. During the same period, commercial publishers took over from older non-profits as the dominant forces in what had, by the late twentieth century, become a highly profitable industry.

There is a long history of governments and other funders supporting research journals. Examples include the UK government (from 1895), the Rockefeller Foundation (from the 1930s), the Nuffield Foundation (in the 1950s) and the US federal government (from the 1960s). But historically, funders were helping non-profit publishers break even, rather than boosting the profits of private enterprise. The success of the profit-making model of scientific publishing in the late twentieth century enabled funders to step back from supporting journals.

The Open Access movement has brought funders back to the table. They are not yet wielding their influence as much as they could, but they need to. Commercial publishers have managed to monetize funder mandates for Open Access (*2*). Author publication fees have become new revenue streams. Rather than democratizing scientific publishing, Open Access has helped commercial publishers generate more profits. More stringent reforms are required to tackle the misaligned drivers of scientific publishing.

Below we describe the drain in more detail and its harms. We argue that publishers' interests have successfully compromised attempts to stop it. We then explain how it may be more effectively addressed. These reforms require research communities to reclaim journals in order to address the drain and Universities, governments and funders to exert their full influence to support them to do so.

**1. The four-fold drain**

*1.1 Money*

Currently, academic publishing is dominated by profit-oriented, multinational companies for whom scientific knowledge is a commodity to be sold back to the academic community who created it. The dominant four are Elsevier, Springer Nature, Wiley and Taylor & Francis, which collectively generated over US$7.1 billion in revenue from journal publishing in 2024 alone, and over US$14 billion in profits between 2019 and 2024 (Table 1A). Their profit margins have always been over 30% in the last five years, and for the largest publisher (Elsevier) always over 37%.

Against many comparators, across many sectors, scientific publishing is one of the most consistently profitable industries (Table S1). These financial arrangements make a substantial difference to science budgets. In 2024, 46% of Elsevier revenues and 53% of Taylor & Francis revenues were generated in North America, charging North American researchers over US$2.27 billion. The National Science Foundation budget that year was $ 9.1 billion and that of NSERC in Canada 1.1 billion.

**Table 1. Scientific Publishing Profits 2019-2024 and APCs 2019-2023**

**Ⓐ · Annual revenues and profits, millions USD, selected publishers**

|  | 2019 | 2020 | 2021 | 2022 | 2023 | 2024 |
|---|---|---|---|---|---|---|
| **Elsevier** | | | | | | |
| Revenue | 3368 | 3456 | 3644 | 3598 | 3809 | 3899 |
| Profit | 1254 | 1311 | 1377 | 1361 | 1449 | 1498 |
| *Profit margin* | *37%* | *38%* | *38%* | *38%* | *38%* | *38%* |
| **Springer Nature** | | | | | | |
| Revenue | 1425 | 1435 | 1598 | 1383 | 1484 | 1530 |
| Profit | 407 | 393 | 455 | 446 | 473 | 489 |
| *Profit margin* | *29%[1]* | *27%[1]* | *28%[1]* | *32%[1]* | *32%* | *32%* |
| **Taylor & Francis** | | | | | | |
| Revenue | 715 | 714 | 750 | 734 | 770 | 892 |
| Profit | 279 | 275 | 281 | 258 | 271 | 327 |
| *Profit margin* | *39%* | *39%* | *37%* | *35%* | *35%* | *37%* |
| **Wiley** | | | | | | |
| Revenue | 939 | 949 | 1015 | 1111 | 1080 | 1043 |
| Profit | 322 | 335 | 357 | 390 | 377 | 331 |
| *Profit margin* | *34%* | *35%* | *35%* | *35%* | *35%* | *32%* |

[1] Springer Nature profits for 2019-22 estimated based on 2023-24 data.
**Source:** Annual financial reports of publishers (see Materials and Methods).

**Ⓑ · Estimated APCs, millions USD, selected publishers**

|  | 2019 | 2020 | 2021 | 2022 | 2023 |
|---|---|---|---|---|---|
| **Elsevier** | 137 | 187 | 322 | 403 | 583 |
| **Frontiers** | 89 | 132 | 230 | 331 | 258 |
| **MDPI** | 168 | 306 | 501 | 657 | 682 |
| **Springer Nature** | 251 | 329 | 429 | 478 | 547 |
| **Wiley** | 101 | 167 | 182 | 255 | 429 |

Source: (3)

Researchers tend to care little about how much is being paid to publishers or feel powerless to affect change. As a result, publishing in this prestige system demonstrates limited price sensitivity. Barriers to entry are still high and the market has become more concentrated. In the past few decades, through mergers and acquisitions, an oligopoly emerged (*4*).

During the transition to digital publishing in the early 2000s, publishers offered libraries so-called "Big Deals" or bundled subscription packages that locked libraries into multi-year, opaque contracts governed by Non-Disclosure Agreements (*5*). Under the Open Access banner, these packages have evolved into "Read-and-Publish Agreements", folding subscriptions and article processing charges (APCs) into a single, similarly opaque invoice for research institutions with little choice but to buy or lose access to scientific literature.

APCs have exacerbated the distortions of commercial publishing. Whereas the Open Access movement aimed to make knowledge freely accessible, publishers found ways to shift paywalls from readers to authors. In some countries in the majority world, where public funds are unavailable, researchers will sometimes meet these costs personally from meagre salaries. APCs now form an increasing part of their lucrative business models (*2*). Between 2019-2023, Haustein and colleagues (*3*) estimated that the top publishers amassed close to $US 9 billion from APC revenues (Table 1B). In today's digital environment, commercial publishers thrive on subscription bundles, APC revenues, and selling the data their publishing work provides (*6*).

The situation is not globally uniform. In Europe, many journals run by learned societies or subject associations are now run in partnership with - or legally transferred to - commercial publishers. However, in Latin America, the majority of journals are still sustained by public universities. As a result, thousands of autonomous, community-led, diamond Open Access journals still thrive outside the publishing oligopoly (*7, 8,* S1). But beyond their sphere, money flowing to commercial publishers drains research resources. Worse still, the incentives publishers face to sell more papers and glean the data they provide drive the other three aspects of the drain we describe.

*1.2 Time*

The number of papers published each year is growing faster than the scientific workforce, with the number of papers per researcher almost doubling between 1996 and 2022 (Figure 1A). This reflects the fact that publishers' commercial desire to publish (sell) more material has aligned well with the competitive prestige culture in which publications help secure jobs, grants, promotions, and awards. To the extent that this growth is driven by a pressure for profit, rather than scholarly imperatives, it distorts the way researchers spend their time.

The publishing system depends on unpaid reviewer labour, estimated to be over 130 million unpaid hours annually in 2020 alone (*9*). Researchers have complained about the demands of peer-review for decades, but the scale of the problem is now worse, with editors reporting widespread difficulties recruiting reviewers. The growth in publications involves not only the authors' time, but that of academic editors and reviewers who are dealing with so many review demands.

Even more seriously, the imperative to produce ever more articles reshapes the nature of scientific inquiry. Evidence across multiple fields shows that more papers result in 'ossification', not new ideas (*10*). It may seem paradoxical that more papers can slow progress until one considers how it affects researchers' time. While rewards remain tied to volume, prestige, and impact of publications, researchers will be nudged away from riskier, local, interdisciplinary, and long-term work. The result is a treadmill of constant activity with limited progress whereas core scholarly practices – such as reading, reflecting and engaging with others' contributions – is de-prioritized. What looks like productivity often masks intellectual exhaustion built on a demoralizing, narrowing scientific vision. Reforms – such as recognizing or compensating peer review, improving evaluation metrics, and rewarding quality over quantity are vital – but they do not address the drive for productivity that stems from for-profit business models.

**Figure 1: Concerning Aspects of Scientific Publishing.**

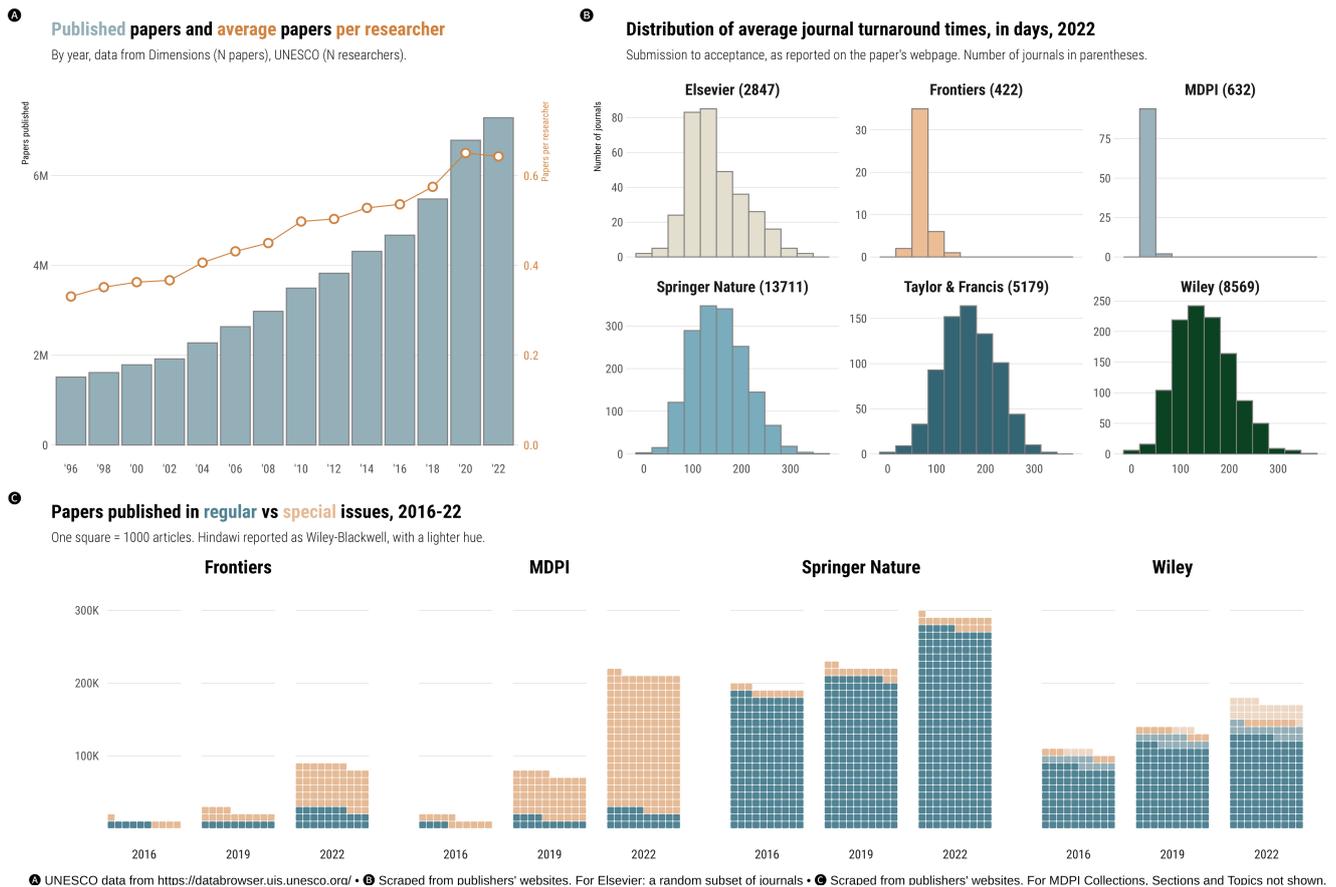

## 1.3 Trust

Maintaining integrity, equity and efficiency within this remarkable growth has become increasingly difficult (*11*). The integrity of scholarly publishing, and its role as a trusted provider of knowledge, rests on authors carrying out research responsibly and writing it up accurately, and on editors and reviewers scrutinising the words, data and images with care. Historically, the space constraints in print-on-paper journals incentivized publishers' and editors' vigilance in this process (if not necessarily guaranteeing quality). The move to digital publishing has removed that constraint and changed publishers' (if not researchers') need for vigilance. Publishers can now make money from publishing quantity as well as quality, whilst still appearing to follow previous conventions of rigorous pre-publication gate keeping.

The proliferation of journals under the Nature brand is one response to this opportunity, as are 'cascade' policies in which publishers seek to place rejected papers in other journals from which they make money. Brands like MDPI, Hindawi (now part of Wiley), Discover (part of Springer Nature) and Frontiers have been particularly adept at seizing on the opportunity presented by a loss of print-on-paper constraints, producing numerous special issues and soliciting authors to contribute to them (Figure 1C and S2).

There is evidence of publishers' new business models affecting editorial independence. Editorial boards have resigned because of the interference of commercial interests, such as 'cascade policies' (S2). The striking lack of variation in turnaround times that some brands achieve across multiple disciplines and hundreds of journals (Figure 1B) suggests a shift from academic editorship to a commercial management of the publishing process.

Such strategies have been seen to backfire, with peer review rings, paper mills, and AI-generated fraud triggering mass retractions and raising concerns about editorial oversight (S2). Some journals have been hijacked (*12*), and entire brands, such as Hindawi, shut down due to the damage caused by paper mills. These are not isolated failures, but symptoms of a research system powerfully shaped by for-profit incentives. The scrutiny and selectivity that peer-reviewed journals were once believed to stand for becomes harder to sustain. Peer review was never flawless, but now the conventional peer review system, based on anonymous and confidential judgements informing decisions to publish (or not), no longer inspires the trust that it once did. The speed and volume of current publishing activity all come at the expense of rigor.

### *1.4 Control*

While academics often retain editorial oversight of journals, even when published by commercial firms, they rarely direct marketing policies, set financial policy, or control brands. Back in the 1960s, a group of UK journal editors had tried to insist that scientific journals should remain under full academic control even when published by commercial firms (*1*). But this proved difficult to implement in an age of increasing demands on academic time, and the increasing complexity of the journal publishing industry.

Yet it is not just the journals themselves that are increasingly controlled by commercial firms: so too are the processes of research evaluation. Some firms are now involved in both journal publishing and compiling the data for quantitative research assessment. Clarivate (now private-equity owned) publishes the Journal Impact Factor via its Journal Citation Reports, a set of prestigious indexes governed by opaque decision-making. Elsevier owns Scopus and an expanding arsenal of analytic platforms that capture the entire research lifecycle. Despite calls for reform (*13*), the continued reliance on these metrics for scientific governance, evaluation and career progression has skewed decision making and incentivized gaming of publication metrics for prestige.

Just as prestige metrics are privately controlled, so too are the institutions tasked with safeguarding publishing integrity – further distancing them from the control of the academic community. The Committee on Publication Ethics (COPE), founded in the 1990s by concerned editors specifically to address misconduct, now includes over 14,000 journals and 100 publishers who act as judge and jury of their own trials.

More broadly, control over scholarly publishing remains concentrated in the Global North with Elsevier, Clarivate, and COPE headquartered in the Netherlands, the US and the UK. This geographic dominance only entrenches Northern norms as the global standard of 'research excellence', thereby marginalizing research published in languages other than English, with regional relevance, or in non-commercial venues (*8*). Platforms such as SciELO, Redalyc, Latindex or African Journals Online offer quality, community-governed alternatives and use clear criteria and local oversight to maintain standards and detect spurious content. Yet, these regional models remain undervalued in evaluation systems shaped by Northern priorities, perpetuating global inequities in defining research quality.

## 2. Stopping the drain

Back in the 1950s, when for-profit journal publishing was just getting going, British scientific leaders predicted that 'the moment commercial gain began to dominate this field, the welfare of the scientific community would suffer' (*1*). They were right – and it is past time that we acted. The first step we must take is to recognize the seriousness of the problems that scientific publishing's engagement with for-profit publishing have become.

The four-fold drain means that for-profit publishing and the prestige economy it sustains no longer serve the interests of the research community. The aims of different actors in scholarly communication remain fundamentally misaligned. This means that we cannot work with commercial publishers to produce system-wide reform. With profit margins above 30%, the temptation will always be to pursue revenue over science.

The futility of working with commercial publishers is clear from the multiple failed attempts to do so. Since the serials crisis emerged in the 1980s (*14*), decades of attempted remedies and over 25 years of Open Access initiatives have applied band-aids to a hemorrhage. Funder policies intended to counteract this have often backfired. For example, in the UK, the system of APC block grants normalized high publishing charges and accelerated the growth of hybrid Open Access. Targets set by Coalition S and others aiming to form fully Open Access journals have not been met. The one constant amidst all these reforms are publishers' profits.

Our analysis also suggests that technological advances will not help. Innovations in AI might improve the efficiency of editorial processes or perhaps assist with some form of preliminary review. This will speed things up. It will generate more papers, and thus more profits for publishers. It will not increase researchers' control over standards. We do not need technology to make commercial publishing quicker. We need to change the structures and incentives governing publishing.

Effective change will thus have to alter the structures – the incentives in, and ownership of, scientific publishing. This will require the interventions of the powerful organizations which shape the evaluation of research and determine its funding. Public funding agencies, foundations and universities have means to act in ways which could transform the current publishing ecosystem. Encouragingly, some are beginning to act more decisively. Some funders already demand all work be preprinted, while others, such as the US NIH, have capped or disallowed the use of funds to pay the APCs of special issue articles (*15*, S2). It is, however, essential that such action should be based on a strategic vision for a future system, something that needs greater discussion and debate.

We propose that scholarly publishing needs to be re-communalized. Universities, libraries, funders and other members of the academic community need to build a system that is community-led and managed, and which works to further research and education. This should involve active support for federated open infrastructures (such as LA Referencia), and investment in community-based publication platforms and non-commercial journals; many of which exist outside the anglo-american space (Érudit, OpenEdition, SciELO). Encouraging innovation in scholarly communication systems, rather than simply reproducing old models is critical, and might include more work in approaches like 'publish, review, curate', which re-conceptualizes the publication and peer review process. Genuine academic oversight needs to be part of this. Policies mandating good practice in disseminating research, such as making work Open Access in affordable ways, are also essential.

The vision of re-communalization may require disruptive change (depending on the political economy of scientific publishing in different world regions) and, as such, cannot be achieved overnight. Major actors in the research environment may work in concert to dismantle the system that currently grants for-profit corporations control over science. However, coordination can be difficult to achieve, and waiting to do so will delay action where it is urgently needed. We, therefore, urge that leading actors in the research community should do just that – lead. Where there are opportunities to act, we recommend that researchers, funders and others take the initiative, and that crucially they do so in ways that further the long-term vision of re-communalization.

Private organisations might provide services to the research community in a future system, but only in ways that do not involve extractive profits. Thus, in the immediate term, decisive intervention in the market should be considered. Competition authorities need to take seriously the dysfunctional nature of the current market and intervene to curb profiteering behaviours and create more meaningful competition, in a manner similar to action that led to the failed merger between Reed Elsevier and Wolters Kluwer almost three decades ago. Radical action by the funders like taking shares in large publishing organizations and exerting pressure from within could also be given consideration – shareholder pressure is a proven lever in changing corporate strategy.

At the same time, a major challenge is to change the incentives and reward structures that shape researchers' publishing behaviours. This means reconfiguring the prestige economy, arguably the most difficult but most crucial task. The current prestige system incentivizes publishing in highly cited journals controlled by commercial publishers. Initiatives like DORA, CoARA, FOLEC and the Barcelona Declaration have contributed to the public discussions on issues of research assessment and reassessing prestige. But they need stronger support from powerful funders, and their advocacy needs more directly to reward research published in community-owned journals and disincentivize publishing in commercially run journals. Again, the funders – governments, universities and foundations - have considerable power here to reward publication outside traditional venues, such as provided by Latin American models, and discourage publishing in commercial outlets.

The costs of inaction are plain: wasted public funds, lost researcher time, compromised scientific integrity and eroded public trust. Today, the system rewards commercial publishers first, and science second. Without bold action from the funders we risk continuing to pour resources into a system that prioritizes profit over the advancement of scientific knowledge.

**Acknowledgments:**
SH thanks Chantal Ripp for assistance with industry profit margins and Margaret Rose for help collecting revenue and profit data of major companies.

**Funding:**
This work contributes to ICTA-UAB "María de Maeztu" Programme for Units of Excellence of the Spanish Ministry of Science and Innovation, CEX2024-001506-M/funded by MICIU/AEI/ 10.13039/501100011033, (DB).

Volkswagen Foundation grant 9C784 "Repercussions of Open Access on Research Assessment (ROARA)" (SH)

**Author contributions:**
Conceptualization: FB, DB, PC, GD, AF, PGB, MAH, SH, VL, CN, SP, JW
Investigation: DB, SH, VL
Visualization: PC
Project administration: DB
Writing: FB, DB, PC, GD, AF, PGB, MAH, SH, VL, CN, SP, JW

**Competing interests:**
Authors declare that they have no competing interests.

**Corrections:**
Version 1 of this paper (released 6[th] Nov 2025) stated that publisher profits were over $12bn between 2019-2024, when they are over $14.7bn. We also under-estimated the NSERC budget.
This second version was released on 17[th] November 2025.

# Drain of Scientific Publishing: Supplementary Materials

**Materials and Methods: Sources for publishers' financial data**

**Wiley.** 'Form 10k' annual reports divide revenues into 'Research', 'Publishing' and 'Solutions'. We included all revenues from 'Research'. We took their 'Adjusted EBITDA' data for this sector.

**Springer-Nature.** Revenues are broken down into four sectors: 'Research', 'Education', 'Health' and 'Professional'. Up until 2022, profits are only given for all four sectors combined. The 2024 report breaks profit down by sector for 2023 and 2024. We took the average contribution of Research in 2023 and 2024, to calculate profits for other years. Figures quoted are the Adjusted Operating Profits.

**Taylor and Francis.** Figures are included in the financial reports of its parent company 'Informa', which include revenues from the book publisher Routledge. Figures quoted are the Adjusted Operating Profit.

**Elsevier.** Annual reports of the parent company Relx include journals as part of their 'Scientific, Technical and Medical' sector as well as the data services that journals provide. We took the 'Adjusted Operating Profits' which are the only profit data provided at the sectoral level. EBITDA data are available, but only for the company as a whole. EBITDA profits tend to be higher than Adjusted Operating Profits. In 2024 for example, the adjusted operating profit for the company as a whole was US$4.095 bn, whereas the EBITDA figure is US$4.767 bn (2024 report page 198). It is possible therefore that the figures we report may under-estimate that drain on academic publishing that publishers' profits constitute.

Wiley reports financials in US dollars, Elsevier and Taylor and Francis in UK Sterling and Springer Nature in Euros. We used currency conversion data, taking the average exchange rate value for each calendar year from this site: www.exchangerates.org.uk. We then converted all data into dollars.

**S1. Examples of alternative publishing practices**

Latin America is a region that highlights for its extensive development of diamond publishing, where a virtous value system is observed as a result of indexation systems that focus on academic quality and independent editorship. This noncommercial and community-led publishing circuit finds its main strength in its public nature and the crucial role of universities. 11,117 active journals are published today through 738 University Portals managed by centralized editorial teams, mostly using OJS. The resilience of diamond publishing is explained by this kind of institutional support. This should be further expanded because many diamond journals are at risk by the constant attempts to buy them by predatory companies and commercial publishers. As the editor of the Latin American Journal of Sedimentology said in an interview: "I receive at least one proposal per month to sell the journal, and the figures offered keep rising" (*16*).

In Tanzania, public universities and research institutions have budgetary allocations specific for supporting journals. Through the Consortium of Tanzania Universities and Research Libraries, 93 academic member institutions across the country have agreed to provide and support open access research information across the country. The consortium supports launching of open access publishing platforms, repositories and the adoption of open access journal policies. The use of institutional repositories has particularly served as an important avenue for dissemination to the research communities, learners, and the general public. These coordinated efforts are new and statistics are not available for measuring the impacts. However, the University of Dar es Salaam (UDSM) - which is the oldest and largest public institution - had its 25 journals in Diamond Open Access until recently in 2024 when two of these entered into contract with Brill. This was locally perceived to be a measure of international recognition and so, perversely, welcome.

Thus, each region has different needs according to its specific context. But the commercial threats remain active everywhere and the prevalence of the dominant concept of "excellence" risks the future of diamond open access.

**S2. Evidence of problems in scientific publishing**
The multiple forms of strain and abuse in scientific publishing are increasingly being systematically documented (*17-19*). Here is a summary of the sources available.

Editorial board resignations. A list of board resignations is kept by Retraction Watch here: https://retractionwatch.com/the-retraction-watch-mass-resignations-list/

Systematic lists of problems can be found on pub-peer: https://pubpeer.com/

Also the group 'United2Act Againt Paper Mills' is also mobilising work against harmful behaviour: https://united2act.org/

There are also commentaries and observations on publishers' behaviour. For example, Springer-Nature launched a new series the 'Discover' series that so closely reassembled MDPI in its purpose and format that it even mimicked the names of MDPI journals. This prompted a humorous response from four of the present authors:

https://the-strain-on-scientific-publishing.github.io/website/posts/discover_nature/

And some of the mistakes are so bad as to be (almost) comical, were not the integrity of science to be at stake. Salient examples include the AI generated image of a massive rats penis (here), and the paper which used capital 'T's instead of error bars (here).

The problems are also visible in the action funders are taking. Gates foundation already demands all work be preprinted (here). The Swiss National Science Foundation have disallowed the use of their funding to pay the APCs of special issue articles (here). The Finnish Publication Forum decided to downgrade hundreds of MDPI and Frontiers journals because of quality concerns (here).

**Supplementary References**

16. F. Beigel, The transformative relation between publishers and editors: Research quality and academic autonomy at stake. *Quantitative Science Studies* **6**, 154–170. (2025).

17. A. Abalkina *et al.*, 'Stamp out paper mills' — science sleuths on how to fight fake research. *Nature* **637**, 1047-1050 (2025).

18. C. Candal-Pedreira *et al.*, Retracted papers originating from paper mills: a cross-sectional analysis of references and citations. *Journal of Clinical Epidemiology* **172**, 111397 (2024).

19. S. J. Porter, L. D. McIntosh, Identifying fabricated networks within authorship-for-sale enterprises. *Scientific Reports* **14**, 29569 (2024).

## Table S1. Comparative Financial Data

**Academic publishers compared to 30 largest companies based on 2024 revenue**
Revenue and profit in million USD, from consolidated revenue. Companies sorted by profit margin.

| | HEADQUARTERS | EMPLOYEES | REVENUE | PROFITS | MARGIN |
|---|---|---|---|---|---|
| **Academic publishing · NAICS 511 · Mean industry net profit margin: 12%** | | | | | Mean: 34% |
| RELX | United Kingdom | 36,400 | 12,057 | 4,088 | 33% |
| └ Elsevier | United Kingdom | 9,700 | 3,899 | 1,497 | 38% |
| Informa | United Kingdom | 11,400 | 4,542 | 1,271 | 27% |
| └ Taylor & Francis | United Kingdom | 11,000 | 892 | 327 | 36% |
| Springer Nature Group | Germany | 9,092 | 1,998 | 554 | 27% |
| └ Springer Nature Research Segment | Germany | 6,125 | 1,529 | 488 | 31% |
| Wiley | United States | 6,400 | 1,042 | 331 | 31% |
| MDPI | Switzerland | 6,650 | -- | -- | -- |
| Frontiers | Switzerland | 1,440 | -- | -- | -- |
| **Information technology · NAICS 334 · Mean industry net profit margin: 27%** | | | | | Mean: 29% |
| Microsoft | United States | 228,000 | 245,122 | 88,136 | 35% |
| Alphabet | United States | 183,323 | 350,018 | 100,118 | 28% |
| Apple | United States | 164,000 | 391,035 | 93,736 | 23% |
| **Oil and gas · NAICS 21111 · Mean industry net profit margin: 21%** | | | | | Mean: 8% |
| Saudi Aramco | Saudi Arabia | 75,118 | 480,446 | 106,246 | 22% |
| ExxonMobil | United States | 60,900 | 349,585 | 35,063 | 10% |
| Chevron | United States | 39,742 | 193,414 | 17,611 | 9% |
| TotalEnergies | France | 102,887 | 241,550 | 18,264 | 7% |
| China National Petroleum Corporation | China | 1,000,800 | 436,875 | 28,677 | 6% |
| Shell | United Kingdom | 96,000 | 289,029 | 16,521 | 5% |
| China Petrochemical Corporation | China | 355,952 | 428,286 | 20,805 | 4% |
| BP | United Kingdom | 100,500 | 194,629 | 6,782 | 3% |
| **Financials · NAICS 52 · Mean industry net profit margin: 19%** | | | | | Mean: 34% |
| Industrial and Commercial Bank of China | China | 415,159 | 109,507 | 51,115 | 46% |
| JPMorgan Chase | United States | 317,233 | 177,556 | 58,471 | 32% |
| Berkshire Hathaway | United States | 392,400 | 371,433 | 88,995 | 23% |
| **Pharmaceuticals · NAICS 3254 · Mean industry net profit margin: 14%** | | | | | Mean: 0% |
| CVS Health | United States | 300,000 | 372,809 | 4,614 | 1% |
| McKesson | United States | 51,000 | 359,051 | 3,481 | 0% |
| Cencora | United States | 46,000 | 293,959 | 1,509 | 0% |
| Cardinal Health | United States | 48,900 | 222,578 | 1,561 | 0% |
| **Energy · NAICS 22111 · Mean industry net profit margin: 14%** | | | | | Mean: 2% |
| State Grid Corporation of China | China | 1,361,423 | 545,954 | 11,465 | 2% |
| **Commodities · NAICS 213112 · Mean industry net profit margin: 13%** | | | | | Mean: 0% |
| Vitol | Switzerland | 1,560 | 331,000 | -- | -- |
| Trafigura | Singapore | 13,086 | 243,201 | 2,758 | 1% |
| Glencore | Switzerland | 83,426 | 230,944 | −2,694 | −1% |
| **Retail · NAICS 44 · Mean industry net profit margin: 7%** | | | | | Mean: 4% |
| Amazon | United States | 1,556,000 | 637,959 | 59,248 | 9% |
| Costco | United States | 333,000 | 254,453 | 7,367 | 2% |
| Walmart | United States | 2,100,000 | 642,637 | 15,511 | 2% |
| **Construction · NAICS 23 · Mean industry net profit margin: 7%** | | | | | Mean: 2% |
| China State Construction Engineering | China | 382,894 | 315,588 | 7,558 | 2% |
| **Automotive · NAICS 336 · Mean industry net profit margin: 6%** | | | | | Mean: 6% |
| Toyota | Japan | 380,793 | 297,628 | 32,636 | 10% |
| Volkswagen Group | Germany | 679,500 | 351,342 | 20,626 | 5% |
| Stellantis | Netherlands | 248,243 | 169,773 | 3,990 | 2% |
| **Health Insurance · NAICS 52411 · Mean industry net profit margin: 3%** | | | | | Mean: 3% |
| UnitedHealth Group | United States | 400,000 | 400,278 | 14,405 | 3% |

Industry net profit margins and industry classification obtained from Dow Jones Factiva Industry Snapshot. List of largest companies obtained from Wikipedia, revenues and profits (in million USD) and number of employees extracted from annual financial reports and converted to USD if necessary.
Source: https://en.wikipedia.org/wiki/List_of_largest_companies_by_revenue
Mean 2024 average exchange rates for USD used: GBP: 1.2781; EUR 1.0822; RMB: 0.1393; JPY: 0.0056; NTD: 0.0312